\input harvmac
%\draftmode
%\def\IR{\relax{\rm I\kern-.18em R}}
%\input epsfDavid

%

\let\includefigures=\iftrue
\let\useblackboard=\iftrue
\newfam\black

%Figure Stuff
\includefigures
\message{If you do not have epsf.tex (to include figures),}
\message{change the option at the top of the tex file.}
\input epsf
\def\figin{\epsfcheck\figin}\def\figins{\epsfcheck\figins}
\def\epsfcheck{\ifx\epsfbox\UnDeFiNeD
\message{(NO epsf.tex, FIGURES WILL BE IGNORED)}
\gdef\figin##1{\vskip2in}\gdef\figins##1{\hskip.5in}% blank space instead
\else\message{(FIGURES WILL BE INCLUDED)}%
\gdef\figin##1{##1}\gdef\figins##1{##1}\fi}
\def\DefWarn#1{}
\def\figinsert{\goodbreak\midinsert}
\def\ifig#1#2#3{\DefWarn#1\xdef#1{fig.~\the\figno}
\writedef{#1\leftbracket fig.\noexpand~\the\figno}%
\figinsert\figin{\centerline{#3}}\medskip\centerline{\vbox{
\baselineskip12pt\advance\hsize by -1truein
\noindent\footnotefont{\bf Fig.~\the\figno:} #2}}
%\bigskip
\endinsert\global\advance\figno by1}
%%%
\else
\def\ifig#1#2#3{\xdef#1{fig.~\the\figno}
\writedef{#1\leftbracket fig.\noexpand~\the\figno}%
%\figinsert\figin{\centerline{#3}}\medskip
%\centerline{\vbox{\baselineskip12pt
%\advance\hsize by -1truein\noindent
%\footnotefont{\bf Fig.~\the\figno:} #2}}
%\bigskip\endinsert
\global\advance\figno by1} \fi

\def\journal#1&#2(#3){\unskip, \sl #1\ \bf #2 \rm(19#3) }
\def\andjournal#1&#2(#3){\sl #1~\bf #2 \rm (19#3) }

\def\ie{{\it i.e.}}
\def\eg{{\it e.g.}}

\noblackbox
%

% Something to deal with sub-sub-sections

\def\unlockat{\catcode`\@=11}
\def\lockat{\catcode`\@=12}

\unlockat
% Something to deal with sub-sub-sections

\def\newsec#1{\global\advance\secno by1\message{(\the\secno. #1)}
\global\subsecno=0\global\subsubsecno=0\eqnres@t\noindent
{\bf\the\secno. #1}
\writetoca{{\secsym} {#1}}\par\nobreak\medskip\nobreak}
\global\newcount\subsecno \global\subsecno=0
\def\subsec#1{\global\advance\subsecno
by1\message{(\secsym\the\subsecno. #1)}
\ifnum\lastpenalty>9000\else\bigbreak\fi\global\subsubsecno=0
\noindent{\it\secsym\the\subsecno. #1}
\writetoca{\string\quad {\secsym\the\subsecno.} {#1}}
\par\nobreak\medskip\nobreak}
\global\newcount\subsubsecno \global\subsubsecno=0
\def\subsubsec#1{\global\advance\subsubsecno by1
\message{(\secsym\the\subsecno.\the\subsubsecno. #1)}
\ifnum\lastpenalty>9000\else\bigbreak\fi
\noindent\quad{\secsym\the\subsecno.\the\subsubsecno.}{#1}
\writetoca{\string\qquad{\secsym\the\subsecno.\the\subsubsecno.}{#1}}
\par\nobreak\medskip\nobreak}

\def\subsubseclab#1{\DefWarn#1\xdef
#1{\noexpand\hyperref{}{subsubsection}%
{\secsym\the\subsecno.\the\subsubsecno}%
{\secsym\the\subsecno.\the\subsubsecno}}%
\writedef{#1\leftbracket#1}\wrlabeL{#1=#1}}% Macros for boxes
\lockat

\def\ie{{\it i.e.}}
\def\eg{{\it e.g.}}

%% MORE MACROS
\def\CM {{\cal M}}
\def\CN {{\cal N}}

\font\manual=manfnt \def\dbend{\lower3.5pt\hbox{\manual\char127}}

\def\IZ{\relax\ifmmode\mathchoice
{\hbox{\cmss Z\kern-.4em Z}}{\hbox{\cmss Z\kern-.4em Z}}
{\lower.9pt\hbox{\cmsss Z\kern-.4em Z}}
{\lower1.2pt\hbox{\cmsss Z\kern-.4em Z}}\else{\cmss Z\kern-.4em
Z}\fi}
\def\half{{1\over 2}}

\def\CM {{\cal M}}
\def\CN {{\cal N}}

% more macros, alphabetically

\def\IZ{\relax\ifmmode\mathchoice
{\hbox{\cmss Z\kern-.4em Z}}{\hbox{\cmss Z\kern-.4em Z}}
{\lower.9pt\hbox{\cmsss Z\kern-.4em Z}}
{\lower1.2pt\hbox{\cmsss Z\kern-.4em Z}}\else{\cmss Z\kern-.4em
Z}\fi}
\def\IB{\relax{\rm I\kern-.18em B}}
\def\IC{{\relax\hbox{$\inbar\kern-.3em{\rm C}$}}}
\def\ID{\relax{\rm I\kern-.18em D}}
\def\IE{\relax{\rm I\kern-.18em E}}
\def\IF{\relax{\rm I\kern-.18em F}}
\def\IG{\relax\hbox{$\inbar\kern-.3em{\rm G}$}}
\def\IGa{\relax\hbox{${\rm I}\kern-.18em\Gamma$}}
\def\IH{\relax{\rm I\kern-.18em H}}
\def\II{\relax{\rm I\kern-.18em I}}
\def\IK{\relax{\rm I\kern-.18em K}}
\def\IP{\relax{\rm I\kern-.18em P}}
\def\IQ{\relax\hbox{$\inbar\kern-.3em{\rm Q}$}}

\def\inbar{\,\vrule height1.5ex width.4pt depth0pt}

\font\cmss=cmss10 \font\cmsss=cmss10 at 7pt
\def\IR{\relax{\rm I\kern-.18em R}}

% Macros for boxes
%
%\def\boxit#1{\vbox{\hrule\hbox{\vrule\kern8pt
%\vbox{\hbox{\kern8pt}\hbox{\vbox{#1}}\hbox{\k
%\hbox{$\displaystyle #1$}\kern8pt}\kern8pt\vrule}\hrule}}}
%
%%% MACROS FOR BOX BOUNDARY CONDS
%%% FROM KAWAI ET AL

\def\makeblankbox#1#2{\hbox{\lower\dp0\vbox{\hidehrule{#1}{#2}%
   \kern -#1% overlap rules
   \hbox to \wd0{\hidevrule{#1}{#2}%
      \raise\ht0\vbox to #1{}% vrule height
      \lower\dp0\vtop to #1{}% vrule depth
      \hfil\hidevrule{#2}{#1}}%
   \kern-#1\hidehrule{#2}{#1}}}%
}%
\def\hidehrule#1#2{\kern-#1\hrule height#1 depth#2 \kern-#2}%
\def\hidevrule#1#2{\kern-#1{\dimen0=#1\advance\dimen0 by #2\vrule
    width\dimen0}\kern-#2}%
\def\openbox{\ht0=1.2mm \dp0=1.2mm \wd0=2.4mm  \raise 2.75pt
\makeblankbox {.25pt} {.25pt}  }

\def\bun#1/#2{\leavevmode
   \kern.1em \raise .5ex \hbox{\the\scriptfont0 #1}%
   \kern-.1em $/$%
   \kern-.15em \lower .25ex \hbox{\the\scriptfont0 #2}%
}

\def\opensquare{\ht0=3.4mm \dp0=3.4mm \wd0=6.8mm  \raise 2.7pt
\makeblankbox {.25pt} {.25pt}  }

%%%%%%%%%%%%%%%%%%%%%%%

\def\sector#1#2{\ {\scriptstyle #1}\hskip 1mm
\mathop{\opensquare}\limits_{\lower 1mm\hbox{$\scriptstyle#2$}}\hskip 1mm}

\def\tsector#1#2{\ {\scriptstyle #1}\hskip 1mm
\mathop{\opensquare}\limits_{\lower 1mm\hbox{$\scriptstyle#2$}}^\sim\hskip 1mm}
%%%
%%%

%% ANOTHER SET OF MACROS

\def\inbar{\,\vrule height1.5ex width.4pt depth0pt}

\font\cmss=cmss10 \font\cmsss=cmss10 at 7pt
\def\IR{\relax{\rm I\kern-.18em R}}

%% new macros

\def\frac#1#2{{#1\over#2}}

\def\half{\frac12}

\def\inbar{\,\vrule height1.5ex width.4pt depth0pt}
\def\IC{\relax\hbox{$\inbar\kern-.3em{\rm C}$}}
\def\IR{\relax{\rm I\kern-.18em R}}
\def\IP{\relax{\rm I\kern-.18em P}}

%
%%%%%%%%%%%%%%%%%%%%%%%%%%%%%%%%%%%%
%
\catcode`\@=11
\def\slash#1{\mathord{\mathpalette\c@ncel{#1}}}
\overfullrule=0pt

\def\EE{{\cal E}}

\def\II{{\cal I}}

\def\LL{{\cal L}}
\def\MM{{\cal M}}
\def\NN{{\cal N}}
\def\OO{{\cal O}}

\def\TT{{\cal T}}

\def\underrel#1\over#2{\mathrel{\mathop{\kern\z@#1}\limits_{#2}}}

\catcode`\@=12

%%%%%%%%%%%%%%%%%%%%%%%%%%%%%%%%%%%%%%%%%%%%%%%%%%%%%%%%%%%%%%

%

\def\exp{{\rm exp}}

%%%%%%%%%%%%%%%%%%%%%%%%%%%%%%%%%%%%%%%%%%%%%%%%%%%%%%%%%%%%%%
% new defs:

%%%%%%%%%%%%%%%%%%%%%%%%%%%%%%%%%%%%%%%%%%%%%%%%%%%%%%%%%%%%%%

\def\frac#1#2{{#1\over#2}}

\def\half{\frac12}

\def\inbar{\,\vrule height1.5ex width.4pt depth0pt}
\def\IC{\relax\hbox{$\inbar\kern-.3em{\rm C}$}}
\def\IR{\relax{\rm I\kern-.18em R}}
\def\IP{\relax{\rm I\kern-.18em P}}

%
%%%%%%%%%%%%%%%%%%%%%%%%%%%%%%%%%%%%
%

%
\catcode`\@=11
\def\slash#1{\mathord{\mathpalette\c@ncel{#1}}}
\overfullrule=0pt

\def\EE{{\cal E}}

\def\II{{\cal I}}

\def\LL{{\cal L}}
\def\MM{{\cal M}}
\def\NN{{\cal N}}
\def\OO{{\cal O}}

\def\TT{{\cal T}}

\def\underrel#1\over#2{\mathrel{\mathop{\kern\z@#1}\limits_{#2}}}

\catcode`\@=12

%%%%%%%%%%%%%%%%%%%%%%%%%%%%%%%%%%%%%%%%%%%%%%%%%%%%%%%%%%%%%%

%

\def\exp{{\rm exp}}

%%%%%%%%%%%%%%%%%%%%%%%%%%%%%%%%%%%%%%%%%%%%%%%%%%%%%%%%%%%%%%
% new defs:

%%%%%%%%%%%%%%%%%%%%%%%%%%%%%%%%%%%%%%%%%%%%%%%%%%%%%%%%%%%%%%%%%%%%%%%%%%%%%%%%%%

%\ZamolodchikovCE
\lref\ZamolodchikovCE{
  A.~B.~Zamolodchikov,
  ``Expectation value of composite field T anti-T in two-dimensional quantum field theory,''
[hep-th/0401146].
%%CITATION = hep-th/0401146%%
}

%\SmirnovLQW
\lref\SmirnovLQW{
  F.~A.~Smirnov and A.~B.~Zamolodchikov,
  ``On space of integrable quantum field theories,''
[arXiv:1608.05499 [hep-th]].
%%CITATION = arXiv:1608.05499%%
}

%\CavagliaODA
\lref\CavagliaODA{
  A.~Cavagli, S.~Negro, I.~M.~Szcsnyi and R.~Tateo,
  ``$T \bar{T}$-deformed 2D Quantum Field Theories,''
[arXiv:1608.05534 [hep-th]].
%%CITAT
}

%\PolchinskiRQ
\lref\PolchinskiRQ{
  J.~Polchinski,
  ``String theory. Vol. 1: An introduction to the bosonic string,''
}

%\GiveonCGS
\lref\GiveonCGS{
  A.~Giveon and D.~Kutasov,
  ``Supersymmetric Renyi entropy in CFT$_{2}$ and AdS$_{3}$,''
JHEP {\bf 1601}, 042 (2016).
[arXiv:1510.08872 [hep-th]].
%%CITATION = arXiv:1510.08872%%
}

%\KutasovXU
\lref\KutasovXU{
  D.~Kutasov and N.~Seiberg,
  ``More comments on string theory on AdS(3),''
JHEP {\bf 9904}, 008 (1999).
[hep-th/9903219].
%%CITATION = hep-th/9903219%%
}

%\GiveonNS
\lref\GiveonNS{
  A.~Giveon, D.~Kutasov and N.~Seiberg,
  ``Comments on string theory on AdS(3),''
Adv.\ Theor.\ Math.\ Phys.\  {\bf 2}, 733 (1998).
[hep-th/9806194].
%%CITATION = hep-th/9806194%%
}

%\GiveonUP
\lref\GiveonUP{
  A.~Giveon and D.~Kutasov,
  ``Notes on AdS(3),''
Nucl.\ Phys.\ B {\bf 621}, 303 (2002).
[hep-th/0106004].
%%CITATION = RI-5-01%%
}

%\MaldacenaKM
\lref\MaldacenaKM{
  J.~M.~Maldacena and H.~Ooguri,
  ``Strings in AdS(3) and the SL(2,R) WZW model. Part 3. Correlation functions,''
Phys.\ Rev.\ D {\bf 65}, 106006 (2002).
[hep-th/0111180].
%%CITATION = CALT-68-2360%%
}

%\IsraelRY
\lref\IsraelRY{
  D.~Israel, C.~Kounnas and M.~P.~Petropoulos,
  ``Superstrings on NS5 backgrounds, deformed AdS(3) and holography,''
JHEP {\bf 0310}, 028 (2003).
[hep-th/0306053].
%%CITATION = hep-th/0306053%%
}

%\GiveonZM
\lref\GiveonZM{
  A.~Giveon, D.~Kutasov and O.~Pelc,
  ``Holography for noncritical superstrings,''
JHEP {\bf 9910}, 035 (1999).
[hep-th/9907178].
%%CITATION = hep-th/9907178%%
}

%\HorowitzEI
\lref\HorowitzEI{
  G.~T.~Horowitz and A.~A.~Tseytlin,
  ``On exact solutions and singularities in string theory,''
Phys.\ Rev.\ D {\bf 50}, 5204 (1994).
[hep-th/9406067].
%%CITATION = hep-th/9406067%%
}

%\ChakrabortySWE
\lref\ChakrabortySWE{
  S.~Chakraborty, A.~Giveon and D.~Kutasov,
  ``$T\bar T$, Black Holes and Negative Strings,''
[arXiv:2006.13249 [hep-th]].
%%CITATION = arXiv:2006.13249%%
}

%\CaselleDRA
\lref\CaselleDRA{
  M.~Caselle, D.~Fioravanti, F.~Gliozzi and R.~Tateo,
  ``Quantisation of the effective string with TBA,''
JHEP {\bf 1307}, 071 (2013).
[arXiv:1305.1278 [hep-th]].
%%CITATION = arXiv:1305.1278%%
}

%\DubovskyWK
\lref\DubovskyWK{
  S.~Dubovsky, R.~Flauger and V.~Gorbenko,
  ``Solving the Simplest Theory of Quantum Gravity,''
JHEP {\bf 1209}, 133 (2012).
[arXiv:1205.6805 [hep-th]].
%%CITATION = arXiv:1205.6805%%
}

%\ChakrabortyNME
\lref\ChakrabortyNME{
  S.~Chakraborty, A.~Giveon and D.~Kutasov,
  ``Comments on D3-Brane Holography,''
[arXiv:2006.14129 [hep-th]].
%%CITATION = arXiv:2006.14129%%
}

%\AtickSI
\lref\AtickSI{
  J.~J.~Atick and E.~Witten,
  ``The Hagedorn Transition and the Number of Degrees of Freedom of String Theory,''
Nucl.\ Phys.\ B {\bf 310}, 291 (1988).
%%CITATION = IASSNS-HEP-88-14%%
}

%\KutasovJP
\lref\KutasovJP{
  D.~Kutasov and D.~A.~Sahakyan,
  ``Comments on the thermodynamics of little string theory,''
JHEP {\bf 0102}, 021 (2001).
[hep-th/0012258].
%%CITATION = hep-th/0012258%%
}

%\GiveonMI
\lref\GiveonMI{
  A.~Giveon, D.~Kutasov, E.~Rabinovici and A.~Sever,
  ``Phases of quantum gravity in AdS(3) and linear dilaton backgrounds,''
Nucl.\ Phys.\ B {\bf 719}, 3 (2005).
[hep-th/0503121].
%%CITATION = hep-th/0503121%%
}

%\CallanAT
\lref\CallanAT{
  C.~G.~Callan, Jr., J.~A.~Harvey and A.~Strominger,
  ``Supersymmetric string solitons,''
In *Trieste 1991, Proceedings, String theory and quantum gravity '91* 208-244 and Chicago Univ. - EFI 91-066 (91/11,rec.Feb.92) 42 p.
[hep-th/9112030].
%%CITATION = hep-th/9112030%%
}

%\SeibergXZ
\lref\SeibergXZ{
  N.~Seiberg and E.~Witten,
  ``The D1 / D5 system and singular CFT,''
JHEP {\bf 9904}, 017 (1999).
[hep-th/9903224].
%%CITATION = hep-th/9903224%%
}

%\AharonyUB
\lref\AharonyUB{
  O.~Aharony, M.~Berkooz, D.~Kutasov and N.~Seiberg,
  ``Linear dilatons, NS five-branes and holography,''
JHEP {\bf 9810}, 004 (1998).
[hep-th/9808149].
%%CITATION = hep-th/9808149%%
}

%\AharonyXN
\lref\AharonyXN{
  O.~Aharony, A.~Giveon and D.~Kutasov,
  ``LSZ in LST,''
Nucl.\ Phys.\ B {\bf 691}, 3 (2004).
[hep-th/0404016].
%%CITATION = hep-th/0404016%%
}

%\SeibergBD
\lref\SeibergBD{
  N.~Seiberg,
  ``Five-dimensional SUSY field theories, nontrivial fixed points and string dynamics,''
Phys.\ Lett.\ B {\bf 388}, 753 (1996).
[hep-th/9608111].
%%CITATION = hep-th/9608111%%
}

%\McGoughLOL
\lref\McGoughLOL{
  L.~McGough, M.~Mezei and H.~Verlinde,
  ``Moving the CFT into the bulk with $T\bar T$,''
[arXiv:1611.03470 [hep-th]].
%%CITATION = arXiv:1611.03470%%
}

%\WittenYR
\lref\WittenYR{
  E.~Witten,
  ``On string theory and black holes,''
Phys.\ Rev.\ D {\bf 44}, 314 (1991).
%%CITATION = IASSNS-HEP-91-12%%
}

%\ArgurioTB
\lref\ArgurioTB{
  R.~Argurio, A.~Giveon and A.~Shomer,
  ``Superstrings on AdS(3) and symmetric products,''
JHEP {\bf 0012}, 003 (2000).
[hep-th/0009242].
%%CITATION = hep-th/0009242%%
}

%\WakimotoGF
\lref\WakimotoGF{
  M.~Wakimoto,
  ``Fock representations of the affine lie algebra A1(1),''
Commun.\ Math.\ Phys.\  {\bf 104}, 605 (1986).
}

%\BernardIY
\lref\BernardIY{
  D.~Bernard and G.~Felder,
  ``Fock Representations and BRST Cohomology in SL(2) Current Algebra,''
Commun.\ Math.\ Phys.\  {\bf 127}, 145 (1990).
%%CITATION = SACLAY-SPH-T-89-113%%
}

%\BershadskyIN
\lref\BershadskyIN{
  M.~Bershadsky and D.~Kutasov,
  ``Comment on gauged WZW theory,''
Phys.\ Lett.\ B {\bf 266}, 345 (1991).
%%CITATION = PUPT-1261%%
}

%\DijkgraafVV
\lref\DijkgraafVV{
  R.~Dijkgraaf, E.~P.~Verlinde and H.~L.~Verlinde,
  ``Matrix string theory,''
Nucl.\ Phys.\ B {\bf 500}, 43 (1997).
[hep-th/9703030].
%%CITATION = hep-th/9703030%%
}

%\GubserKV
\lref\GubserKV{
  S.~S.~Gubser, A.~Hashimoto, I.~R.~Klebanov and M.~Krasnitz,
  ``Scalar absorption and the breaking of the world volume conformal
invariance,''
Nucl.\ Phys.\ B {\bf 526}, 393 (1998).
[hep-th/9803023].
%%CITATION = hep-th/9803023%%
}

%\IntriligatorAI
\lref\IntriligatorAI{
  K.~A.~Intriligator,
  ``Maximally supersymmetric RG flows and AdS duality,''
Nucl.\ Phys.\ B {\bf 580}, 99 (2000).
[hep-th/9909082].
%%CITATION = hep-th/9909082%%
}

%\HyunJV
\lref\HyunJV{
  S.~Hyun,
  ``U duality between three-dimensional and higher dimensional black holes,''
J.\ Korean Phys.\ Soc.\  {\bf 33}, S532 (1998).
[hep-th/9704005].
%%CITATION = hep-th/9704005%%
}

%\BerkoozUG
\lref\BerkoozUG{
  M.~Berkooz, A.~Sever and A.~Shomer,
  ``'Double trace' deformations, boundary conditions and space-time singularities,''
JHEP {\bf 0205}, 034 (2002).
[hep-th/0112264].
%%CITATION = hep-th/0112264%%
}

%\WittenUA
\lref\WittenUA{
  E.~Witten,
  ``Multitrace operators, boundary conditions, and AdS / CFT correspondence,''
[hep-th/0112258].
%%CITATION = hep-th/0112258%%
}

%\KastorEF
\lref\KastorEF{
  D.~A.~Kastor, E.~J.~Martinec and S.~H.~Shenker,
  ``RG Flow in N=1 Discrete Series,''
Nucl.\ Phys.\ B {\bf 316}, 590 (1989).
%%CITATION = EFI-88-31-CHICAGO%%
}

%\ForsteWP
\lref\ForsteWP{
  S.~Forste,
  ``A Truly marginal deformation of SL(2, R) in a null direction,''
Phys.\ Lett.\ B {\bf 338}, 36 (1994).
[hep-th/9407198].
%%CITATION = hep-th/9407198%%
}

%\ZamolodchikovVX
\lref\ZamolodchikovVX{
  A.~B.~Zamolodchikov,
  ``From tricritical Ising to critical Ising by thermodynamic Bethe ansatz,''
Nucl.\ Phys.\ B {\bf 358}, 524 (1991).
%%CITATION = ENS-LPS-327-1991%%
}

%\HertogRZ
\lref\HertogRZ{
  T.~Hertog and G.~T.~Horowitz,
  ``Towards a big crunch dual,''
JHEP {\bf 0407}, 073 (2004).
[hep-th/0406134].
%%CITATION = hep-th/0406134%%
}

%\HertogHU
\lref\HertogHU{
  T.~Hertog and G.~T.~Horowitz,
  ``Holographic description of AdS cosmologies,''
JHEP {\bf 0504}, 005 (2005).
[hep-th/0503071].
%%CITATION = hep-th/0503071%%
}

%\ElitzurKZ
\lref\ElitzurKZ{
  S.~Elitzur, A.~Giveon, M.~Porrati and E.~Rabinovici,
  ``Multitrace deformations of vector and adjoint theories and their holographic duals,''
JHEP {\bf 0602}, 006 (2006).
[hep-th/0511061].
%%CITATION = hep-th/0511061%%
}

%\CrapsCH
\lref\CrapsCH{
  B.~Craps, T.~Hertog and N.~Turok,
  ``On the Quantum Resolution of Cosmological Singularities using AdS/CFT,''
Phys.\ Rev.\ D {\bf 86}, 043513 (2012).
[arXiv:0712.4180 [hep-th]].
%%CITATION = arXiv:0712.4180%%
}

%%%%%%%%%%%%%%%%%%%%%%%%%%%%%%%%%%%%%%%%%%%%%%%%%%%%%%%%%%%%%%%%%%%%%%%%%%%%%
%\rightline{....}
\Title{
%\rightline{hep-th/yymmnnn}
} {\vbox{
\bigskip\centerline{$T\bar T$ and LST}}}
\medskip
\centerline{\it Amit Giveon${}^{1}$, Nissan Itzhaki${}^{2}$ and David Kutasov${}^{3}$}
\bigskip
\smallskip
\centerline{${}^{1}$Racah Institute of Physics, The Hebrew
University} \centerline{Jerusalem 91904, Israel}
\smallskip
\centerline{${}^{2}$ Physics Department, Tel-Aviv University, Israel} \centerline{Ramat-Aviv, 69978, Israel}
\smallskip
\centerline{${}^3$EFI and Department of Physics, University of
Chicago} \centerline{5640 S. Ellis Av., Chicago, IL 60637, USA }

\bigskip\bigskip\bigskip
\noindent

It was recently shown that the theory obtained by deforming a general two dimensional conformal theory by the irrelevant operator $T\bar T$ is solvable. In the context of holography, a large class of such theories can be obtained by studying string theory on $AdS_3$. We show that a certain single trace $T\bar T$ deformation of the boundary $CFT_2$ gives
rise in the bulk to string theory in a background that interpolates between $AdS_3$ in the IR and a linear dilaton spacetime in the UV, \ie\ to a two dimensional vacuum of Little String Theory. This construction provides holographic duals for a large class of vacua of string theory in asymptotically linear dilaton spacetimes, and sheds light on the UV behavior of $T\bar T$ deformed $CFT_2$. It may provide a step towards holography in flat spacetime.

\vglue .3cm
%\vskip 2cm
\bigskip

\Date{1/17}

\newsec{Introduction}

Vacua of quantum gravity in asymptotically anti de Sitter spacetimes are holographically dual to Quantum Field Theories (QFT's) which approach a renormalization group fixed point (a Conformal Field Theory) in the UV. This is reflected in the fact that the entropy of the bulk theory, which is dominated at high energies by large $AdS$ black holes, grows as $E\to\infty$ like $E^\alpha$, with $\alpha={d-2\over d-1}<1$ (for $AdS_d$), in agreement with the growth of the entropy in a $d-1$ dimensional CFT.

One of the most important open problems in quantum gravity is to generalize the highly successful AdS/CFT paradigm to other spacetimes, such as flat Minkowski spacetime $\IR^{d-1,1}$. There are many indications that holography plays a central role in such spacetimes as well, but there is no useful description of the boundary theory. One way to see the difficulty is to note that the entropy of very massive black holes in $\IR^{d-1,1}$ grows like $E^\alpha$ with $\alpha={d-2\over d-3}>1$, so the dual theory must exhibit this growth as well. This implies that it cannot be a local QFT in the usual sense (an RG flow connecting two fixed points).

It has been known for a long time that there is an interesting intermediate case, Little String Theory (LST). From the bulk perspective, it corresponds to spacetimes of the (asymptotic) form $\IR_\phi\times \IR^{d-1,1}$ where $\IR_\phi$ is labeled by a radial coordinate $\phi$, and has the property that the dilaton depends linearly on $\phi$, at least near the boundary at $\phi\to\infty$, where the string coupling goes to zero. Such backgrounds arise naturally in string theory near $NS5$-branes \CallanAT, and singularities of Calabi-Yau manifolds \GiveonZM, and are believed to exhibit holography as well \AharonyUB.

The black holes that govern the high energy thermodynamics of LST have an entropy that grows like the energy. In this sense, LST can be thought of as an intermediate case between anti de Sitter spacetime, in which the entropy grows like $E^\alpha$ with $\alpha<1$, and flat Minkowski spacetime, where $\alpha>1$. There are other ways in which this is the case. For example, both linear dilaton and anti de Sitter spacetimes are solutions to (dilaton) gravity with negative cosmological constant, while in flat spacetime the cosmological constant vanishes. On the other hand,  the time it takes signals to propagate from the bulk to the boundary is finite in $AdS$ spacetime and infinite in the other two.

All of the above suggests that understanding holography in asymptotically linear dilaton spacetimes might provide a useful step towards holography in flat spacetime. What makes this particularly interesting is that the Hagedorn behavior of the entropy at high energies implies that the dual cannot be a standard local QFT, like in flat spacetime, and it would be nice to understand it microscopically.

Most of the work on LST in the past treated it as a bulk theory in an asymptotically linear dilaton spacetime, and there was no useful definition of the dual, or boundary, theory. The aim of this note is to take advantage of recent progress in field theory  \refs{\SmirnovLQW,\CavagliaODA}  to (partially) rectify this situation. We will argue that a large class of two dimensional conformal field theories deformed by a certain single trace $T\bar T$ operator provide a boundary description of two dimensional vacua of LST. In particular, they give a microscopic understanding of the Hagedorn entropy of black holes in asymptotically linear dilaton backgrounds of the form $\IR_\phi\times \IR_t\times S^1$.

The plan of this note is the following. In section 2 we discuss some of the results of \refs{\SmirnovLQW,\CavagliaODA}. In particular, we point out that the spectrum of a $CFT_2$ on a spatial circle deformed by the irrelevant operator $T\bar T$, interpolates between that of a $CFT_2$ at low energies and a spectrum with Hagedorn growth at high energies. In section 3, as preparation for a discussion of the $T\bar T$ deformation in holography, we review some elements of string theory on $AdS_3$, and the LST that underlies it.  In particular, we present a geometry that gives rise to a bulk realization of an RG flow from a theory with Hagedorn density of states in the UV to a $CFT_2$ in the IR, which is qualitatively similar to that appearing in $T\bar T$ deformed $CFT_2$.

In section 4 we discuss the $T\bar T$ deformation of $CFT_2$ in the context of holography. We show that a certain natural single trace $T\bar T$  deformation corresponds in the bulk to a marginal current-current deformation of the worldsheet theory on $AdS_3$. We investigate this deformation in section 5, and find that it corresponds to string theory in the bulk geometry that interpolates between $AdS_3$ in the IR and a linear dilaton spacetime in the UV discussed in section 3. Thus, it describes a particular (two dimensional) vacuum of LST. In section 6 we compare the high energy thermodynamics of the boundary theory (a $T\bar T$ deformed $CFT_2$) to that of the bulk theory, obtained by studying black holes in the deformed background, and find agreement between the two. In section 7 we comment on our results.

A note on terminology: when discussing the (deformed) $CFT_2$ in the context of holography, we will alternate between referring to it as a boundary (as opposed to bulk) and a spacetime (as opposed to worldsheet) theory.

\newsec{$T\bar T$ deformation of $CFT_2$}

In this section we discuss some aspects of the recent work \refs{\SmirnovLQW,\CavagliaODA} (see also \refs{\DubovskyWK,\CaselleDRA}) on the theory obtained by deforming the Lagrangian of a two dimensional conformal field theory $(CFT_2)$ by
\eqn\deltal{\delta\LL=tT\bar T,}
where $T$ and $\bar T$ are the holomorphic and anti-holomorphic components of the stress tensor, respectively, and the composite operator $T\bar T$ is defined at finite coupling $t$ as \ZamolodchikovCE
\eqn\ccc{T\bar T(y)=\lim_{x\to y} \left(T(x)\bar T(y)-\Theta(x)\Theta(y)\right).
}
$\Theta$ is the trace of the stress tensor, related to $T$ and $\bar T$ by the conservation equations
\eqn\bbb{\partial_{\bar x} T=\partial_x\Theta;\qquad \partial_x\bar T=\partial_{\bar x}\Theta.}
The operator $T\bar T$ has dimension four. Hence the coupling $t$ in \deltal\ has units of length squared; in particular, it is irrelevant (in the RG sense). The resulting theory approaches the original CFT at long distances, whereas at short distances one in general expects the theory to lose predictive power.

The authors of  \refs{\SmirnovLQW,\CavagliaODA} used techniques from integrable field theory to study the deformation \deltal. In particular, they calculated exactly the spectrum of the theory on a circle of circumference $R$ and found it to be
\eqn\spectll{E(R,t)=-{R\over 2t}+\sqrt{{R^2\over 4t^2}+{4\pi\over t}(h-{c\over 24})}~,}
where $c$ is the central charge of the original CFT.  The coupling $t$ must be taken to be positive for the theory to have a vacuum \SmirnovLQW.

In the infrared limit $t\to 0$, \spectll\ approaches the standard CFT result,
\eqn\spectcft{E(R,0)={4\pi\over R} (h-{c\over 24}),}
corresponding to a state\foot{In \spectll\ we exhibited the result for states with $h=\bar h$. The generalization to the case $h\not=\bar h$ is known.} with $L_0=\bar L_0=h$. Equation \spectll\ describes the change of the energy of such a state as we turn on the perturbation. A few comments about \spectll\ are in order at this point:
\item{(1)} This equation relates the three dimensionless quantities that appear in the problem,
\eqn\defquant{b={4\pi t\over R^2},\qquad
{\cal E}={ER\over 2\pi},\qquad
M= 2h-{c\over 12}~.}
In terms of these quantities, it takes the form
\eqn\ebm{{\cal E}(b,M)=-{1\over b}+\sqrt{{1\over b^2}+{2M\over b}}~.}
\item{(2)} $b$ in \defquant\ can be thought of as the value of the coupling $t$ \deltal\ at the scale $R$. Small $b$ corresponds to weak coupling at distances $\ge R$.  In this weak coupling regime, there is a large number of states, those for which $|M|\ll 1/b$, whose energies are little changed by the perturbation and are given to a good approximation by \spectcft, $\EE\simeq M$. As $|Mb|$ grows, the deviations of the spectrum \ebm\ from the CFT one increase. In particular, for high energies $(Mb\gg1$) one finds
\eqn\largeee{\EE\simeq\sqrt{2M\over b},\;\; {\rm or} \;\; E\simeq\sqrt{2\pi M\over t}~.
}
Interestingly, the spectrum becomes $R$ independent in this limit; the energy scale is set by the coupling $t$. Note that in the weak coupling regime $b\ll1$, the energy scale  in \largeee, $1/\sqrt t$, is much higher than the energy scale of the original CFT, which is of order $1/R$.
\item{(3)} The coupling plays an important role for low lying states above the $SL(2,\IR)$ invariant ground state as well. Consider, for example, the ground state itself, which corresponds to $h=0$, $M=-c/12$, \defquant. As explained in the previous comment, the properties of this state start to deviate significantly from the original CFT when $bc$ becomes of order one. In particular, for $bc>6$ one finds that the energy of the ground state \ebm\ becomes complex. For $c\sim 1$ this does not happen at weak coupling, but for large $c$ it does. One can think of $bc$ as a `t Hooft coupling for this theory. We will mostly discuss the regime of weak coupling $(b\ll 1)$ and arbitrary `t Hooft coupling $bc$.
\item{(4)} One can use the energy formula \ebm\ to analyze the entropy of the deformed theory. Since this formula relates each state in the theory \deltal\  to a state in the original CFT, the entropy $S(\EE)$ is given by the usual Cardy form, $S=S_c(M)$, with $M$ expressed in terms of $\EE$ via \ebm. The Cardy entropy of the CFT is
\eqn\cardyf{S_c(M)\simeq2\pi\sqrt{{c\over3}M}~,}
valid for $M\gg1$. Using \ebm, this can be written as
\eqn\entpert{S(\EE)=2\pi\sqrt{{c\over 6}\left(2\EE+b\EE^2\right)}~,}
which is expected to be valid for $\EE\gg1$.  For $1\ll\EE \ll1/b$, this reduces to the original CFT entropy \cardyf. In the opposite regime, $\EE \gg1/b$ one finds instead
\eqn\hagedorn{S\simeq 2\pi\sqrt{bc\over 6}\EE=\sqrt{2\pi ct\over 3} E~.
}
This is a Hagedorn entropy, $S=\beta_H E$, with inverse Hagedorn temperature
\eqn\betahh{\beta_H=\sqrt{2\pi ct\over 3}~.}
Note that $\beta_H$ is equal to the circumference of the circle, $R$, at the point where the ground state energy becomes complex  (corresponding to $bc=6$ in \defquant). This is a UV/IR relation reminiscent of the relation between the mass of the winding tachyon and the Hagedorn temperature in (perturbative) critical string theory (see \eg\ \AtickSI). Like there, in models with fermions the state with $M=-c/12$ corresponds to fermions with anti-periodic boundary conditions around the circle.
\item{(5)} The fact that the $T\bar T$ deformed theory \deltal\ has a Hagedorn density of states implies that its UV behavior is not governed by an RG fixed point. A theory with this behavior is not expected to have local operators, such as the stress tensor \bbb. Thus, some of the assumptions that were used in \ZamolodchikovCE\ are not satisfied in this case. Nevertheless, the results of \refs{\SmirnovLQW,\CavagliaODA} suggest that the analysis of \ZamolodchikovCE\ is still valid.
\item{(6)} There are RG flows in QFT that look in the vicinity of an IR fixed point as $T\bar T$ deformations of a $CFT_2$, but lead in the UV to a fixed point rather than to a Hagedorn spectrum (see \eg\ \refs{\KastorEF,\ZamolodchikovVX}). The reason for the difference with the results of \refs{\SmirnovLQW,\CavagliaODA} is that in the latter case the theory looks like a $T\bar T$ deformation at all scales, and not just in the IR. 

\noindent
We see that at high energies the theory \deltal\ does not approach a UV fixed point, but rather has a Hagedorn spectrum \hagedorn. There is a class of interacting non-gravitational theories that is known to have a Hagedorn density of states at high energies -- Little String Theory. Thus, it is natural to ask whether the theories \deltal\ give rise to two dimensional vacua of LST. In the remainder of this note we will argue that in a large class of examples this is indeed the case.

\newsec{Aspects of the $AdS_3/CFT_2$ correspondence}

To study the results of \refs{\SmirnovLQW,\CavagliaODA}  in the context of the AdS/CFT correspondence, we start in this section with a brief review of some relevant aspects of string theory on $AdS_3$ (see \eg\ \refs{\GiveonNS,\KutasovXU} for more detailed discussions). We will take the $AdS_3$ background to be supported by a flux of the Neveu-Schwartz $B$-field. In this case the worldsheet theory on $AdS_3$ is solvable -- it is described by the WZW model on the $SL(2,\IR)$ group manifold. The worldsheet dynamics is governed to a large extent by a left-moving $SL(2,\IR)$ current algebra generated by the currents $J^a(z)$, $a=3,+,-$, and another, right-moving, current algebra $\bar J^a(\bar z)$, both at level $k$. The level determines the size of the $AdS$ space in string units, $R_{AdS}=\sqrt{k} l_s$.

A typical vacuum of string theory on $AdS_3$ takes the form
\eqn\adsvac{AdS_3\times \NN,}
where $\NN$ is a compact CFT, whose central charge is determined by the condition that the total worldsheet central charge of \adsvac\ is critical (twenty six for the bosonic string, and fifteen for the superstring). A well studied example is superstring theory with $\NN=S^3\times T^4$, which describes the near-horizon geometry of $k$ $NS5$-branes wrapped on $T^4\times S^1$ and $p$ fundamental strings wrapped on $S^1$. The $S^3$ corresponds to the angular directions in the $\IR^4$ transverse to both the strings and the fivebranes, and is described by a WZW model on the $SU(2)$ group manifold. The total levels of $SU(2)$ and $SL(2,R)$ are both equal to $k$.

The worldsheet $SL(2,\IR)_L\times SL(2,\IR)_R$ currents play an important role in the study of the theory. The zero modes of these currents give conserved global charges in the spacetime theory, which correspond to the global part of the conformal group of the spacetime (or boundary) theory. In particular, $J^3$ gives $L_0$, while $J^\pm$ give $L_{\pm1}$, and similarly for the other worldsheet and spacetime chirality. The full (anti) holomorphic stress tensor $T(x)$ $(\bar T(\bar x))$ of the spacetime theory is constructed in terms of these currents and other observables on $AdS_3$ \KutasovXU.

An observation \GiveonZM\ that will play an important role in our discussion below is that vacua of the form \adsvac\ are closely related to two dimensional vacua of LST. Indeed, consider string theory in the background
\eqn\twodlst{\IR_t\times \IR_\phi\times S^1\times \NN,}
where the first factor is time, the second a spatial direction with linear dilaton along it, and the rest as before. The slope of the linear dilaton, $Q$, is determined by the criticality condition.  As $\phi\to\infty$, the string coupling $g_s\simeq\exp(-Q\phi/2)$ goes to zero. $\phi\to\infty$ is the boundary of the spacetime \twodlst, on which the dual theory lives. As $\phi\to-\infty$ the coupling diverges. To study physics in this region requires some understanding of the strong coupling behavior of the theory.

String theory in the background \twodlst\ is believed to be holographically dual to a non-local, non-gravitational theory \AharonyUB. The high energy behavior of the entropy is dominated by black holes, which are obtained by replacing $\IR_t\times \IR_\phi$  in \twodlst\ by the two dimensional black hole background. This black hole is described by an exactly solvable worldsheet theory, which corresponds to the coset $SL(2,\IR)/U(1)$  \WittenYR. The high energy behavior of the entropy is Hagedorn, $S(E)=\beta_H E$, with
\eqn\betahhh{\beta_H=2\pi\sqrt{k} l_s.}
Some thermodynamic properties of this black hole  were studied in \refs{\KutasovJP,\GiveonMI}.

The relation between \adsvac\ and \twodlst\ can be thought of as follows \GiveonZM. One starts with the background \twodlst\ and adds to it $p$ fundamental strings wrapping the $S^1$. The addition of the strings modifies the background in the IR (\ie\ in the region $\phi\to-\infty$). In particular, the string coupling no longer grows without bound there; rather it saturates at a value $g_s^2\sim 1/p$. Thus, for large $p$ the coupling is small everywhere, and one can study bulk string theory in this background using perturbative techniques.

The full background takes in this case the form \GiveonZM\ (see also \IsraelRY)
\eqn\mmthree{\MM_3\times\NN,}
where $\NN$ is as in \adsvac, \twodlst, and  the three dimensional background $\MM_3$ is
\eqn\interpolate{\eqalign{
ds^2=&f_1^{-1}l_s^2d\gamma d\bar\gamma+kl_s^2d\phi^2,\cr
e^{2\Phi}=&{v\over p}e^{-2\phi}f_1^{-1},\cr
dB=&2i e^{-2\phi}f_1^{-1}\epsilon_3,}}
where $l_s\gamma=x^1+x^0$, $l_s\bar\gamma=x^1-x^0$,
\eqn\somedefs{f_1=1+ {1\over k}e^{-2\phi},}
and $v$ is a constant associated with the compact CFT $\NN$. Recall that $x^1$ is periodically identified, $x^1\sim x^1+R$.

The background \mmthree, \interpolate\ interpolates between the LST background \twodlst\ near the boundary at $\phi\to\infty$ and the $AdS_3$ background \adsvac\ in the infrared $\phi\to-\infty$. It describes an RG flow from a non-local theory with a Hagedorn spectrum in the UV, to a standard $CFT_2$ in the IR. The transition between the two occurs at a value of $\phi$ determined by the requirement that the two terms in $f_1$ \somedefs\ are comparable. We can choose this scale to take any convenient value, and this has effectively been done in \somedefs. The RG flow \interpolate\ will be the focus of our discussion below.

In the example mentioned above, where $\NN=S^3\times T^4$, the UV geometry \twodlst\ is the CHS geometry \CallanAT\ of $k$ $NS5$-branes wrapped on $T^4\times S^1$, and the non-local theory in question is the six dimensional LST of $k$ fivebranes compactified on a torus. The RG flow \interpolate\ interpolates in this case between this non-local theory in the UV and the $CFT_2$ obtained by studying the system of $k$ $NS5$-branes and $p$ fundamental strings in the infrared limit, which is dual to string theory on $AdS_3\times S^3\times T^4$.

Another observation which will play a role below concerns the structure of the boundary CFT corresponding to the background \adsvac. It has been proposed  \refs{\ArgurioTB,\GiveonCGS} that at large $p$ this theory has the form of a symmetric product
\eqn\mmpp{\MM^p/S_p~,}
where $\MM$ is a CFT with central charge $c_\MM=6k$. Roughly speaking, $\MM$ can be thought of as the CFT associated with a single string added to the background \twodlst, and the structure \mmpp\ relies on the fact that at large $p$ the interaction between the $p$ strings in the background \twodlst\ goes to zero. The status of \mmpp\ is not completely clear, but we will assume it below. Our results provide some further support for this assumption.

Since for large $p$ string theory in the background \adsvac\ is weakly coupled, one can use perturbative (worldsheet) techniques to study it. In particular, one can construct vertex operators that describe low lying local operators in the boundary theory \refs{\GiveonNS,\KutasovXU}. In terms of \mmpp\ one can think of these vertex operators as ``single trace'' operators in the symmetric orbifold CFT.  More precisely, they correspond to operators of the form
\eqn\pcopies{\sum_{i=1}^p\OO_i(x),}
where $\OO_i$ correspond to a particular operator $\OO(x)\in\MM$ living in the $i$'th factor in \mmpp, with the sum over $i$ imposing $S_p$ invariance.

\newsec{The $T\bar T$ deformation in $AdS_3$}

In order to make contact with the discussion of \refs{\SmirnovLQW,\CavagliaODA}, we need to construct the $T\bar T$ deformation in string theory on $AdS_3$.
One possibility is to use the vertex operator of the stress tensor in \KutasovXU, whose construction we will review shortly, and consider the double trace deformation by the product of the vertex operators for $T$ and $\bar T$. We will next argue that there is also a single trace $T\bar T$ deformation, which is easier to study in the bulk.

To understand it, it is useful to start with a brief review of the construction of the stress tensor in \KutasovXU.\foot{For simplicity, we will discuss the construction in the bosonic string. The generalization to the superstring is explained in \KutasovXU.} There are two observables that play a key role in this construction. One is the current
\eqn\jjxxzz{J(x;z)=2xJ_3(z)-J^+(z)-x^2 J^-(z),}
which combines the $SL(2,\IR)_L$ currents into a single object labeled by the auxiliary variable $x$. This variable can be thought of as position on the boundary of $AdS_3$. There is also a right-moving analog of \jjxxzz\ with the opposite worldsheet and spacetime chirality $\bar J(\bar x;\bar z)$.

The second observable is
\eqn\defphih{\Phi_h(x;z)={1\over\pi}\left(1\over|\gamma-x|^2e^\phi+e^{-\phi}\right)^{2h}~,
}
which is an eigenfunction of the Laplacian on $AdS_3$, and gives rise in the quantum theory to a primary of the worldsheet and spacetime Virasoro, with worldsheet dimension $\Delta_h=\bar\Delta_h=-h(h-1)/(k-2)$ and spacetime dimension $(h,h)$.

In terms of the operators \jjxxzz, \defphih, the vertex operator of the spacetime stress-tensor is given by
\eqn\stresst{T(x)={1\over 2k}\int d^2z(\partial_x J\partial_x\Phi_1+2\partial_x^2J\Phi_1)\bar J(\bar x;\bar z).}
As explained in \KutasovXU, this operator is physical (\ie\ BRST invariant), its spacetime scaling dimension is $(2,0)$, it is holomorphic, $\partial_{\bar x}T=0$ (as expected from the unitarity of the spacetime CFT), and it satisfies the standard OPE algebra of the holomorphic stress-tensor in the spacetime $CFT_2$.  The anti-holomorphic stress tensor $\bar T(\bar x)$ is constructed similarly by flipping all the chiralities in \stresst, $(x,z,J)\leftrightarrow(\bar x,\bar z,\bar J)$.

The $T\bar T$ deformation of \refs{\SmirnovLQW,\CavagliaODA} corresponds in terms of the above discussion to adding to the worldsheet action the product of the vertex operators for $T$ and $\bar T$. Since each of these vertex operators is given by an integral over the worldsheet, this leads to a non-local worldsheet deformation, typical of double trace deformations in holography.

To construct the single trace deformation we will be interested in, we observe (following \KutasovXU) that there is another vertex operator we can consider, that combines elements of the constructions of $T$ \stresst\ and $\bar T$. That vertex operator is
\eqn\ddxx{D(x)=\int d^2z(\partial_x J\partial_x+2\partial_x^2J)(\partial_{\bar x} \bar J\partial_{\bar x}+2\partial_{\bar x}^2\bar J)\Phi_1.
}
Since the left-moving part of the vertex operator \ddxx\ is the same as that of $T(x)$ \stresst, while its right-moving part is the same as that of $\bar T$, we know that from the point of view of the spacetime Virasoro algebra it is a quasi-primary operator of dimension $(2,2)$. It is natural to ask what is the spacetime interpretation of this vertex operator.

To answer this question it is useful to look back at the discussion of the previous section, and in particular to recall that the spacetime CFT takes (at large $p$) the symmetric product form \mmpp. In terms of this structure, the stress tensor vertex operator \stresst\ corresponds to the sum of the stress tensors of the different factors in \mmpp,
\eqn\ttotal{T(x)=\sum_{i=1}^p T_i(x).}
It is a special case of \pcopies, with the operator $\OO(x)$ being the stress tensor of the CFT $\MM$.

In terms of the structure \mmpp\ there is a natural conjecture for the interpretation of the operator $D(x)$ \ddxx\ in the spacetime CFT,
\eqn\dddxxx{D(x)=A\sum_{i=1}^p T_i(x)\bar T_i(\bar x),}
which again has the form \pcopies, with $\OO(x)\sim T(x)\bar T(\bar x)$, the product of the holomorphic and anti-holomorphic stress tensors in $\MM$. The constant $A$ can be determined by computing the OPE of $T(x)$ \stresst\ with $D(y)$ \ddxx.  From the spacetime CFT we expect to get
\eqn\singope{T(x) D(y)={Ac_\MM/2\over (x-y)^4}\bar T(\bar y)+\cdots.}
On the worldsheet, this calculation can be done using the results of \KutasovXU. In addition to determining the constant $A$, the structure of this OPE provides evidence for the identification \dddxxx.

Further evidence for this identification can be obtained by calculating the OPE of $D(x)$ with other operators in the spacetime CFT. A large class of such operators is obtained by taking a Virasoro primary operator $V(z)$ with dimension $\Delta_V=\bar\Delta_V$ in $\CN$ \adsvac, and dressing it with an operator from the $AdS_3$ CFT, $\Phi_h$,
\eqn\ooxx{\OO(x)=\int d^2z \Phi_h(x;z) V(z).}
The operator \ooxx\ is physical as long as its total worldsheet dimension is one, $\Delta_V+\Delta_h=1$.  From the perspective of the boundary theory, it corresponds to a combination of the form \pcopies. Again, one can calculate the OPE $D(x)\OO(y)$ in the CFT and in the $AdS$ and compare. We leave the details of the worldsheet calculation of this OPE as well as of \singope\ to another publication.

The operator \dddxxx\ is interesting from the perspective of studying $T\bar T$ deformations of a CFT. Adding it to the Lagrangian of the spacetime CFT \mmpp\ corresponds to deforming each copy of $\CM$ by a $T\bar T$ deformation with the same coupling $t$. It leads to a spacetime theory of the form
\eqn\deformedmm{\left(\MM_t\right)^p/S_p~,}
where $\MM_t$ is the CFT $\MM$ deformed by a $T\bar T$ deformation \deltal, and we can use the results of \refs{\SmirnovLQW,\CavagliaODA} to study it. Thus, it is interesting to understand what happens to the bulk string theory when we add to the worldsheet Lagrangian the vertex operator $\int d^2x D(x)$, with $D(x)$ given by \ddxx. We have
\eqn\finaldd{\int d^2x D(x,\bar x)=\int d^2x \int d^2z\partial_x^2 J\partial_{\bar x}^2\bar J \Phi_1=4\int d^2x \int d^2z J^-(z)\bar J^-(\bar z)\Phi_1(x;z).}
In the first step we integrated by parts, using the fact that the boundary terms at infinity vanish due to the rapid drop-off of the wave function $\Phi_1$ \defphih\ there. In the second step we used the definition of $J(x;z)$ \jjxxzz. In principle  we have to be careful about defining the composite operator $J^-(z)\Phi_1(x;z)$ since there is a short distance singularity when $J^-$ approaches $\Phi_1$, but since the relevant OPE is
\eqn\opejminus{J^-(z)\Phi_h(x;w)\simeq-{\partial_x\Phi_h\over z-w}~,}
the residue of the worldsheet pole is a total derivative in spacetime, and the integral over $x$ in \finaldd\ kills this singular term.

We can now perform the $x$ integral in \finaldd. Since the only quantity that depends on $x$ is $\Phi_1$, the integral we need to perform is $\int d^2x\Phi_1(x;z)$. Plugging in the form of $\Phi_1$ \defphih, we find an elementary integral, which gives a constant. Finally, we have
\eqn\ddworldsheet{\int d^2x D(x,\bar x)\simeq \int d^2z J^-(z)\bar J^-(\bar z),}
where we did not keep track of an overall multiplicative constant.

We conclude that the $T\bar T$ deformation of the CFT $\MM$ in \mmpp\ is described from the bulk perspective by a $J^-\bar J^-$ deformation of the $SL(2,\IR)$ CFT. In the next section we will discuss this deformation of the bulk theory.

\newsec{The $J^-\bar J^-$ deformation of $AdS_3$}

In the previous section we argued that a particular $T\bar T$ deformation \dddxxx\ of the boundary CFT \mmpp\ corresponding to an $AdS_3$ background in string theory is described in the bulk by a deformation of the worldsheet theory by a term of the form
\eqn\jminus{\delta\CL_{\rm ws}=\lambda J^-(z)\bar J^-(\bar z).}
In this section we will discuss some properties of this deformation.

The first notable property of the deformation \jminus\ is that it is truly marginal on the worldsheet. It might seem surprising at first glance that the irrelevant $T\bar T$ deformation of the spacetime theory corresponds on the worldsheet to a marginal one, but this is exactly what we should expect. The fact that the worldsheet deformation is marginal is the statement  that we have a solution of the bulk string theory equations of motion for all values of $\lambda$. The bulk background dual to the $T\bar T$ deformed theory \deformedmm\ should of course have this property.

The irrelevant nature of the $T\bar T$ deformation must be reflected in the fact that the effect of the deformation \jminus\ should increase in the UV, \ie\ as we approach the boundary of $AdS_3$. As a simple first check, one can calculate the (spacetime) scaling dimension of the coupling $\lambda$ in the boundary CFT. As mentioned earlier, the operators $J^-$ and $\bar J^-$ give rise to the spacetime Virasoro generators $L_{-1}$ and $\bar L_{-1}$, respectively. Thus, the operator $J^-\bar J^-$ has spacetime scaling dimension $(1,1)$, and the corresponding coupling, $\lambda$ has scaling dimension $(-1,-1)$, in agreement with that of the coupling $t$  in \deltal. Hence, we expect this coupling to decrease in the infrared, and increase in the ultraviolet in the spacetime theory.

One way to see the effects of $\lambda$ on the bulk geometry is to use the Wakimoto variables \WakimotoGF\ (see also \refs{\BernardIY,\BershadskyIN}). In terms of these variables, the $SL(2,\IR)$ sigma model is described by the worldsheet Lagrangian
\eqn\wakimoto{\CL_{\rm ws}=\partial\phi\bar\partial\phi-{2\over\alpha_+}\widehat R\phi+\beta\bar\partial\gamma+\bar\beta\partial\bar\gamma-
\beta\bar\beta\exp\left(-{2\over\alpha_+}\phi\right),
}
where $\alpha_+^2=2k-4$.

The Lagrangian \wakimoto\ looks superficially like an asymptotically linear dilaton spacetime near the boundary at $\phi\to\infty$, but this is in fact misleading. Indeed, integrating out the fields $\beta$, $\bar\beta$, that appear quadratically in the Lagrangian, one gets back the $AdS_3$ sigma model in the Poincare coordinates $(\phi,\gamma,\bar\gamma)$, and the linear dilaton disappears when performing the Gaussian path integral.

In the description \wakimoto\ the currents $J^-$, $\bar J^-$ take a particularly simple form, $J^-(z)=\beta(z)$ and $\bar J^-(\bar z)=\bar\beta(\bar z)$. Thus, the perturbation \jminus\ corresponds in this language to adding to the worldsheet Lagrangian a term proportional to $\beta\bar\beta$. The coefficient of this term can be set to $\pm 1$ by using the symmetry of \wakimoto\ corresponding to opposite rescaling of $(\beta,\bar\beta)$ and $(\gamma,\bar\gamma)$ combined with a shift of $\phi$.

For one sign, the deformed worldsheet Lagrangian takes the form
\eqn\newwak{\CL_{\rm ws}=\partial\phi\bar\partial\phi-{2\over\alpha_+}\widehat R\phi+\beta\bar\partial\gamma+\bar\beta\partial\bar\gamma-
\beta\bar\beta\left[1+\exp\left(-{2\over\alpha_+}\phi\right)\right].
}
This Lagrangian exhibits a bulk realization of the RG flow of the boundary theory. As $\phi\to-\infty$ (\ie\ in the infrared), the correction to the Lagrangian due to the $J^-\bar J^-$ deformation decreases, and the  deformed Lagrangian \newwak\ approaches the undeformed one \wakimoto.  On the other hand, as $\phi\to\infty$ the correction becomes important. In particular, the coefficient of $\beta\bar\beta$ in \newwak\ now approaches a constant, rather than going exponentially to zero, as it did in \wakimoto. Thus, when we integrate out $\beta$, $\bar\beta$, we get a sigma model for $(\phi,\gamma,\bar\gamma)$ that asymptotes to a linear dilaton spacetime, rather than to $AdS_3$. In fact, performing the path integral over $\beta$, $\bar\beta$, we find\foot{To be precise, in the bosonic string this is true for large $k$, while in the superstring it is true exactly, in terms of a suitably defined level $k$.} the background $\CM_3$ discussed in section 3, equations \interpolate, \somedefs, \refs{\ForsteWP,\GiveonZM}.

We arrive at the following picture for the role of the $T\bar T$ deformation \ddworldsheet\ in string theory on $AdS_3$. As reviewed in section 3, it is very natural to view a string background of the form $AdS_3\times\NN$ \adsvac\ as obtained  from a two dimensional vacuum of LST, which corresponds to a linear dilaton background $\IR_t\times\IR_\phi\times S^1\times\NN$ \twodlst, upon addition of $p$ fundamental strings wrapped around the $S^1$. The full background \interpolate, \somedefs\ describes an RG flow, which approaches in the infrared the fixed point corresponding to the $AdS_3$ background \adsvac, which is a CFT of the form \mmpp, and in the UV describes a theory with a Hagedorn spectrum.

The $T\bar T$ deformation \dddxxx\ of the infrared CFT \mmpp\ provides the boundary description of this RG flow. This description involves a flow up the RG, which normally is considered to be problematic in QFT. But here, the authors of \refs{\SmirnovLQW,\CavagliaODA} argue that this leads to a well defined, predictive structure, presumably due to the symmetries preserved by the $T\bar T$ deformations. This is in agreement with the bulk perspective, where the deformation in question corresponds to a marginal worldsheet perturbation, which is expected to be well behaved.

We finish this section with a few comments:

\item{(1)} A byproduct of our analysis concerns holography for asymptotically linear dilaton spacetimes. As mentioned in the introduction, in the past LST was defined as the bulk theory in an asymptotically linear dilaton spacetime \AharonyUB, and there was no useful independent description of the dual, or boundary, theory. Our results combined with those of \refs{\SmirnovLQW,\CavagliaODA} lead to such a description. To the extent that the $T\bar T$ deformed CFT can be studied using conventional field theoretic techniques, it provides an independent description of the bulk theory on asymptotically linear dilaton spacetimes of the form \interpolate, \somedefs. Conversely, our construction clarifies the nature of the UV behavior of $T\bar T$ deformed $CFT_2$. The authors of \SmirnovLQW\ pointed out that this theory is non-local. From our perspective, this non-locality is due to the fact that the UV theory is a LST, which is inherently non-local in the UV.

\item{(2)} In our analysis above we implicitly assumed that the sign of the $J^-\bar J^-$ deformation (and thus of the associated $T\bar T$ deformation) is such that the coefficient of $\beta\bar\beta$ in \newwak\ is $1+\exp\left(-{2\over\alpha_+}\phi\right)$. For the opposite sign, this coefficient would have been $-1+\exp\left(-{2\over\alpha_+}\phi\right)$, which would lead in \interpolate, \somedefs\ to $f_1=-1+ {1\over k}e^{-2\phi}$. The resulting background has a singularity at the place where $f_1$ vanishes, and the large positive $\phi$ region, where the boundary of the spacetime is supposed to be, is hidden behind it. This background seems to be unphysical. In the boundary theory, the ``good'' sign corresponds to positive $t$ in \spectll, and the ``bad'' one to $t<0$. The authors of \SmirnovLQW\ claim that the $T\bar T$ deformed theory does not have a vacuum for $t<0$. This is in agreement with our picture, where the sickness is the singularity at finite $\phi$.\foot{The fate of singularities in the context of gauge/gravity duality is discussed \eg\ in 
\refs{\HertogRZ\HertogHU\ElitzurKZ-\CrapsCH}.} In both pictures it is the UV part of the theory that is cut off. In the boundary theory this is the statement that for $t<0$ the energies \spectll\ become complex for large $h$; in the bulk, this is due to the singularity that prevents one from going to arbitrarily large positive $\phi$.\foot{A recent paper \McGoughLOL\ considered the regime of negative $t$ in the context of holography, for a different $T\bar T$ deformation -- the double trace deformation \refs{\WittenUA, \BerkoozUG} corresponding to the total $T$ and $\bar T$. The picture proposed in that paper involves an $AdS_3$ geometry with a Dirichlet wall  at a finite radial position.  It would be interesting to understand the relation of that work to ours, where the $AdS_3$ geometry is deformed, and the role of the wall is played by the  singularity described above.}

\item{(3)} $AdS_3$ vacua in string theory can be studied using the powerful techniques based on $SL(2,\IR)$ current algebra. While the $J^-\bar J^-$ deformation breaks the $SL(2,\IR)_L\times SL(2,\IR)_R$ current algebra, the worldsheet theory is still amenable to an algebraic analysis, since it corresponds to a coset CFT of the form $SL(2,\IR)\times U(1)/U(1)$ \HorowitzEI.

\newsec{Spectrum and thermodynamics}

In this section we use the holographic duality described in the previous sections to provide a microscopic description of the Bekenstein-Hawking entropy of black holes in asymptotically linear dilaton backgrounds. To do this, we compare the entropy of highly excited states in the $T\bar T$ deformed CFT \deformedmm\ to the entropy of the corresponding black holes.

We start on the boundary side of the duality. A generic state of energy $E$ in the theory \deformedmm\ is obtained by dividing the energy between the $p$ factors of the $T\bar T$ deformed $CFT_2$ $\MM_t$,
\eqn\totaleee{E=\sum_{j=1}^p E_j.}
The total entropy of such states is
\eqn\totentr{S_{\rm tot}(E)=\sum_j S_\MM(E_j).}
For comparison with the bulk, we would like to take the total energy $E$ to be large, and maximize \totentr\ over all partitions of the energy \totaleee, subject to the constraint that the total energy is fixed.

It is easy to see that a partition of the energy with all $E_j=E/p$ is a stationary point of \totentr. That point is a local maximum of the entropy if and only if the entropy of $\MM_t$, $S_\MM(E)$, satisfies the constraint
\eqn\poscpecheat{{\partial^2S_\MM\over\partial E^2}<0~,}
\ie\ the system has positive specific heat. In our case, assuming that all the $E_j$ are large, the microscopic entropy is given, in terms of dimensionless energy $\EE$, by \entpert,
\eqn\entmmtt{S_\MM(\EE)=2\pi\sqrt{{c_\MM\over 6}\left(2\EE+b\EE^2\right)}~.}
Recall that $c_\MM=6k$, the central charge of the undeformed $CFT_2$, $\MM$, and \entmmtt\ is valid for $\EE\gg 1$.

The first derivative of \entmmtt\ gives the (dimensionless) inverse temperature,
\eqn\invtemp{\beta(\EE)=S_\MM'(\EE)=2\pi\sqrt{c_\MM\over6}(1+b\EE)(2\EE+b\EE^2)^{-\half}~.
}
The second derivative \poscpecheat\ is
\eqn\betaprime{\beta'(\EE)=-2\pi\sqrt{c_\MM\over6}{1\over(2\EE+b\EE^2)^{3\over2}}
}
We see that $\MM_t$ is a system with positive specific heat -- increasing the energy increases the temperature.\foot{It is interesting that without the strings, the specific heat of LST is negative, due to quantum string corrections to the leading, Hagedorn, thermodynamics \KutasovJP. That effect is present here as well, but it is subleading in the $1/p$ expansion, for energies that scale like $p$.} Hence, the maximal entropy configuration in \deformedmm\ is one  in which all $E_j$ are equal, and one has
\eqn\t{S(\EE)=pS_\MM(\EE/p)=2\pi\sqrt{2kp \EE +kb\EE^2 },}
where we used the fact that $c_\MM=6k$. This result is valid for $\EE\gg p$, and interpolates between the undeformed CFT behavior, which is obtained for $p\ll \EE\ll p/b$, and the UV Hagedorn behavior for $\EE\gg p/b$.

We will next compare the result \t\ to the thermodynamic entropy of black holes in the bulk geometry \interpolate, \somedefs.
The black holes in question are described by the metric and dilaton  \HyunJV
\eqn\interpolatebh{\eqalign{
&ds^2=-{f_{\EE} \over f_1} (dx^0)^2+{1\over f_1} (dx^1)^2+ {f_5 \over f_{\EE}} dr^2 , \cr
&e^{-2(\Phi -\Phi_0)}={f_1 \over f_5}, }}
with
\eqn\somedefsbh{f_1=1+ {r_1^2 \over r^2},\;\;\; f_{\EE}=1-{r_0^2 \over r^2},\;\;\; f_5={k l_s^2 \over r^2}.}
$r_0$ and $r_1$ in \somedefsbh\ depend on the mass of the black hole; see \ChakrabortySWE\ for a detailed discussion.
There is also a $B$ field that we did not write, and $x^1 \sim x^1 + R$.

In the background \interpolatebh\ there is a competition between $f_{\EE}$ and $f_1$. For a low mass black hole, $r_0\ll r_1$, the horizon of the black hole is deep inside the $AdS_3$ region, and the solution \interpolatebh\ is given by the BTZ black hole one to a good approximation. Hence, the entropy is given by the Cardy entropy with $c=6kp$. For large black hole mass, $r_0\gg r_1$, the horizon is deep inside the linear dilaton region, the solution is given to a good approximation by the $SL(2,\IR)/U(1)$ black hole times $S^1$, and the entropy exhibits Hagedorn growth; see \ChakrabortySWE.

This behavior is similar to that found in the field theory analysis \t\ (see the discussion around \entpert). There, the transition between the $CFT_2$ and LST behavior of the entropy occurred at $\EE\sim p/b\sim pR^2/t$. In the gravity calculation it occurs at $r_0\sim r_1$, which means \ChakrabortySWE, $\EE\sim pR^2/l_s^2$. Thus, the two calculations agree if we set the coupling $t$ to $t\sim l_s^2$.

To make this more precise, we compute the entropy of the black hole \interpolatebh\ using the Bekenstein-Hawking formula. We get \ChakrabortySWE
\eqn\bhentropy{S_{BH}=  2\pi \sqrt{{4k\pi^2 l_s^2 \EE^2\over R^2}+2pk \EE},}
which agrees with \t\ for
\eqn\formttttt{t=\pi l_s^2 .}
Thus, we see that in the bulk theory the dimensionless coupling $b$ \defquant\ takes the form $b=4\pi^2 l_s^2/R^2$. The condition $b\ll 1$ implies that the radius of the $x^1$ circle is much larger than $l_s$. Note also that plugging in the value of $t$ \formttttt\ and $c=6k$ into \betahh\ leads to the correct Hagedorn temperature \betahhh.

The discussion of this section extends the standard agreement between field theoretic and gravitational entropies from the case that the energy above extremality is small both compared to that of the fivebranes and that of the strings,\foot{$E\ll pR/2\pi l_s^2$ or, equivalently  \defquant, \formttttt, $\EE\ll p/b$.} to the case where the energy is much smaller than that of the fivebranes but not necessarily that of the strings. On the gravity side, the horizon of the black hole \interpolatebh\ is always in the near-horizon geometry of the fivebranes, but not necessarily in that of the strings.

The value of the deformation parameter $t$ \formttttt\ leads to an interesting way of writing the spectrum of the $T\bar T$ deformed theory $\MM_t$ \spectll,
\eqn\nnneee{
\left(E+R\TT\right)^2=(R\TT)^2+{4\over\alpha'}\left(h-{c_\MM\over 24}\right),
}
where we used the usual definition for $\alpha'=l_s^2$, and the string tension $\TT=1/2\pi\alpha'$. Equation \nnneee\ looks like the standard formula for the energy spectrum of a free string winding around a circle of circumference $R$, with the transverse space described by the CFT $\MM$. The factor $h-(c_\MM/24)$ (with $\bar h=h$), is the left and right-moving excitation level of the string, measured relative to the stretching energy $R\TT$.

The result \nnneee\ has a natural interpretation from the perspective of our construction. We saw that the $T\bar T$ deformed theory \deformedmm\ is obtained by adding $p$ fundamental strings wrapped on $S^1$ to a linear dilaton background corresponding to a two dimensional vacuum of LST \twodlst. If we think of this background as the near-horizon geometry of a system of $NS5$-branes wrapped around some manifold, as in \GiveonZM,  the resulting background \interpolate\ describes strings confined to the worldvolume of the fivebranes. The worldsheet theory of these strings in unitary gauge is a CFT $\MM$, which depends on the precise fivebrane system. The spectrum \nnneee\ describes excitations of these strings, which are nothing but the eponymous strings of LST.

As an example of the above discussion, consider the case where $\NN$ in \twodlst\ is $S^3\times T^4$, corresponding to $k$ $NS5$-branes wrapped around $T^4\times S^1$ \CallanAT. In this case the CFT $\MM$ can be thought of as the target space of a string bound to the collection of $k$ fivebranes. The spectrum \nnneee\ describes  excitations of such a string. If we want the fermions on the string to have periodic boundary conditions, we have to study the theory $\MM$ in the Ramond sector for both the left and right-movers. The ground state energy in that sector corresponds to $h=\bar h=c_\MM/24$, and the last term in \nnneee\ is the  excitation level above the ground state.

\newsec{Discussion}

In this note we argued that recent progress in field theory \refs{\SmirnovLQW,\CavagliaODA} sheds new light on an old problem in string theory, concerning holography in asymptotically linear dilaton spacetimes. Such theories give rise to vacua of Little String Theory, a non-local non-gravitational theory with a Hagedorn high energy density of states. We focused on the specific case of two dimensional LST, and studied the sector of the theory with $p\gg1$ fundamental strings. The infrared limit of the theory is in this case a two dimensional conformal field theory \mmpp, which has a well studied $AdS_3$ dual \adsvac.  We proposed that the holographic dual of string theory in the asymptotically linear dilaton spacetime \twodlst,  is the $T\bar T$ deformed $CFT_2$ \deformedmm.
 
One consequence of our analysis is a microscopic understanding of the Hagedorn entropy of two dimensional black holes in asymptotically linear dilaton spacetime. The microstates of these black holes are states in the $T\bar T$ deformed $CFT_2$  \deformedmm, whose entropy agrees with that of the corresponding black holes.

The boundary theories that figure in our discussion (and that of \refs{\SmirnovLQW,\CavagliaODA}) do not fall into the usual paradigm of quantum field theory. Mathematically well defined QFT's correspond to RG flows that connect a UV fixed point to an IR one. In our case, we start from an IR fixed point and flow up the RG to a theory that is non-local at short distances. Usually, attempting to flow up the RG leads to loss of predictive power, associated with the fact that there are in general multiple RG trajectories that look in the infrared like the fixed point theory perturbed by a particular irrelevant operator.

From that perspective, it may seem puzzling why the theories studied in \refs{\SmirnovLQW,\CavagliaODA} and here are well behaved and predictive. In the field theory analysis the answer seems to be the extra symmetries effectively imposed on the theory by the requirement of integrability, which lead to a particular RG trajectory. There are examples in QFT, especially with a large amount of supersymmetry, where analogous flows up the RG lead to sensible results, such as \SeibergBD, and of course critical string theory itself, where a low energy gauge theory with irrelevant interaction is completed to a sensible theory valid for arbitrarily high energies. In all these cases, it seems that from the perspective of the low energy theory, the choice of RG trajectory is determined by a large symmetry.

In our string theory discussion, the flow up the RG in spacetime seems to be sensible since the corresponding worldsheet deformation, \jminus, is in fact marginal. This looks superficially different from the field theory reasoning, but the duality suggests that the two are related. In particular, one should be able to identify the conserved charges that underlie the analysis of  \refs{\SmirnovLQW,\CavagliaODA} in string theory with the $J^-\bar J^-$ deformation, which would make the relation manifest.

An important question regarding the work of \refs{\SmirnovLQW,\CavagliaODA} is whether it provides a complete description of a unitary quantum theory. These papers address the question what happens to the energies of states in a $CFT_2$ after a $T\bar T$ deformation.  One can ask whether unitarity requires the presence of additional states that are not captured by this analysis, \eg\ are there states that disappear in the infrared limit. In our string theoretic description, one can for example ask whether the field theoretic analysis captures  the continuum of states living in the infinite throat $\IR_\phi$ in the bulk description. These states are known to be important in the study of off-shell Greens' functions of LST  \AharonyXN, and they give an important contribution to the thermodynamics of the theory \KutasovJP. We leave the study of these states, and in particular their relation to long strings in $AdS_3$ \SeibergXZ, to future work.

It is natural to ask whether there is an analog of our picture in higher dimensions. In particular, one can study $\CN=4$ SYM perturbed by the irrelevant operator corresponding to adding back the $1$ in the harmonic function of the threebranes (see \eg\  \refs{\GubserKV,\IntriligatorAI}). If this deformation preserves integrability, one may be able to use it to study holography in flat spacetime; see \ChakrabortyNME\  for a recent discussion.

\bigskip\bigskip
\noindent{\bf Acknowledgements:}
We thank O. Aharony, M. Berkooz, S. Datta, S. Elitzur and E. Martinec for discussions.
The work of AG and NI is supported in part by the I-CORE Program of the Planning and Budgeting Committee and the Israel Science Foundation (Center No. 1937/12), and by a center of excellence supported by the Israel Science Foundation (grant number 1989/14). DK is supported in part by DOE grant DE FG02-13ER41958. DK thanks Tel Aviv University and the Hebrew University for hospitality during part of this work.

\listrefs

\end